\documentclass{du-journals}

\journal{Iranian Journal of Astronomy and Astrophysics}

\title{Study of Solar Magnetic and Gravitational Energies Through the Virial Theorem}
\subtitle{Solar Energies and Virial Theorem}

\author[1]{Elham Bazyar}
\address[1]{Research Institute for Astron. \& Astrophys. of Maragha (RIAAM), 55134-441 Maraghe, Iran; email: bazyar$_{-}$84@yahoo.com}

\author[2]{Ali Ajabshirizadeh}
\address[2]{Research Institute for Astron. \& Astrophys. of Maragha (RIAAM), 55134-441 Maraghe, Iran; email: a$_{-}$adjab@tabrizu.ac.ir}

\author[3]{Zahra Fazel}
\address[3]{Astrophysics Depart., Physics Faculty, University of Tabriz, Tabriz, Iran; email: z$_{-}$fazel@tabrizu.ac.ir}

\author[4]{Jean-Pierre Rozelot}
\address[4]{Nice university, OCA-CNRS-Lagrange Depart., Av. Copernic, 06130 Grasse, France; email: rozelot@obs-azur.fr}

\begin{document}

\begin{abstract}
Virial theorem is important for understanding stellar structures. It produces an interesting connection between the
magnetic energy and the gravitational one. Using the general form of the virial theorem including the magnetic field
(toroidal magnetic field), we may explain the solar dynamo model in related to variations of the magnetic and
gravitational energies. We emphasize the role of the gravitational energy in sub-surface layers which has been
certainly minored up to now. We also consider two types of solar outer shape (spherical and spheroidal) to study
the behavior of magnetic and gravitational energies. The magnetic energy affects the solar shape, while the
gravitational energy is not changed by the considered shapes of the Sun.
\end{abstract}


\section{Introduction}

There is a general agreement that the magnetic field of the Sun is generated by the dynamo process in which the kinetic energy is converted into the magnetic energy. The solar magnetic field may be divided into the strong and the weak components. Sunspots are the most prominent manifestation of the strong component. They show a regular cyclic behavior and they obey Hale's polarity rules. The strong field is assumed to result from dynamo action in the overshooting convection layer at the bottom of the convective zone. Differential rotation in the convective zone builds up a large toroidal magnetic field which is stored in the sub-adiabatically stratified medium. By contrast, the magnetic buoyancy instability is set up for producing strong fields of about $10^{5}$$G$ \cite{Moreno92,Schus94,Fer93,Fer95}. The weak field is irregular; it can be due to the turbulent dynamo action in the upper convective zone. In a weak regime the instability provides a dynamo \cite{Schmitt04,Fer94}, which regenerates the poloidal magnetic field and therefore close the dynamo cycle. A strong instability leads to the rise of the flux tubes through the convective zone, producing the bipolar active regions at the solar surface \cite{Cal95}. The azimuthal magnetic field that results from winding-up of an initially weak field is subject to instabilities.
\\In this paper we study the importance of the virial theorem to understand the role of the gravitational energy in the sub-surface layers in comparison with magnetic energy. At first section, virial theorem is presented. In section 2, the solar shape is discussed. Section 3 gives the calculation used in this work. In section 4, results are presented. Finally, discussion and conclusion are followed in section 5.

\section{Virial Theorem}

Virial theorem can be important for understanding the structure and the evolution of stars.
A general form of the virial theorem including the magnetic field was given by \cite{Cha53}:
\begin{equation}
\frac{1}{2}\frac{d^2J}{dt^2}=2K + \Omega + M + 2\int PdV +S
\end{equation}
\begin{equation}
J= \int\rho r^2 dV,
M= \int \frac{B^2}{2\mu_{0}}dV,
K= \frac{1}{2}\int \rho v^2 dV,
\Omega= \frac{1}{2}\int \rho \Psi dV
\end{equation}
where $J$ is the inertial moment of the mass distribution in the region of radius $R$;
$K$, the kinetic energy of (macroscopic) mass motion; $\Omega$, the gravitational potential energy
(with $\Psi$ being the gravitational potential); $M$, the magnetic energy; $P$, the gas pressure and $S$,
a surface integral over the boundary $\partial R$ (with normal vector, $n$) of the region $R$, viz.
\begin{equation}
S= -\oint P(r\cdot n)ds + \frac{1}{\mu_{0}}\oint (r\cdot B)(B\cdot n)ds,
\end{equation}
here, $P_{tot} = P_{gas}+ \frac{B^2}{2\mu_{0}}$ is the total pressure (gas and magnetic). In the case of an ideal gas with
constant ratio of specific heats, $\gamma=\frac{C_{p}}{C_{v}}$, the pressure integral over the region $R$ can be related
to the thermal (or internal) energy, $U$, that is
\begin{equation}
\int PdV = (\gamma -1)U.
\end{equation}
The region of integration  can be the whole or a part of the studied star.
If the stellar structure is in equilibrium, then it follows the virial theorem that
\begin{equation}
3(\gamma -1)U +M +U =0
\end{equation}
On the other hand, the total energy, $E$, is given by $E=U+M+\Omega$. Now by eliminating $U$:
\begin{equation}
E=- \frac{3\gamma -4}{3(\gamma-1)}(|\Omega|-M).
\end{equation}
Considering Eq. (7), a necessary condition is obtained for the dynamical stability of an equilibrium state, that is
\begin{equation}
(3\gamma -4)(|\Omega|-M) > 0,
\end{equation}
which indicates a relation between the magnetic and the gravitational energies. Hence, study of variations of the
solar gravitational energy will be significant. It can be the key to identify the seat region (the leptocline, \cite{Lef09})
of many phenomena, for example: an oscillation phase of the seismic radius together with a non-monotonic expansion of this
radius with depth, a change in the turbulent pressure, an inversion in the radial gradient of the rotation velocity rate at
about 50° in latitude, opacity changes and superadiabaticity. It could be also the cradle of hydrogen and helium ionization
processes and probably the seat of in-situ magnetic fields and luminosity production. In addition, the solar radius changes
as identified through space missions \cite{Emi00,Kuhn04,Pip11}, or through temporal
dependence of the f-modes are certainly induced both by magnetic and gravitational variations during
the course of solar activity \cite{Pap98}.
The mean equilibrium state of a star, at any given time, will follow the reduced magnetic virial theorem, for
equilibrium states:
\begin{equation}
2K +\Omega +M =0
\end{equation}
where the radiant energy content is ignored, as being comparatively small. Moreover, the total energy must be conserved,
for all states:
\begin{equation}
K + \Omega +M =C.
\end{equation}
where $C$ is a constant. Note that the radiant energy losses at the stellar surface are not included, because the considered
time intervals are short enough with respect to the star's cooling time (Kelvin-Helmholtz time, i.e. the time needed to
radiate away a significant fraction of the thermal energy of a star). Following \cite{Stot06}, when $M=0$, hence $2K_{0} + \Omega_{0}=0$
and $K_{0} + \Omega_{0} =C$. If the internal changes are not dynamically fast, then $K=K_{0}$ and $\Omega - \Omega_{0} =-M$.
This implies that the magnetic energy which decays into Joule heat ultimately gets converted into the gravitational potential energy.
We differentiate Eqs. (9) and (10) for any change in the magnetic energy, $\delta M$, then
\begin{equation}
\delta K =0, \delta M= -\delta \Omega.
\end{equation}
Radiant energy losses at the surface can be neglected, as they are of second order. From $\Omega = -\int \frac{GM(r)}{r} dM(r)$,
we deduce $\frac{\delta\Omega}{\Omega} = -\zeta \frac{\delta R}{R}$, where $0<\zeta\leq 1$ (in the case of a homogenous structure
of a star, $\zeta =1$). Then
\begin{equation}
\frac{\delta M}{|\Omega|} = -\zeta \frac{\delta R}{R},
\end{equation}
where the sign is important. These equations are very simply based on the virial theorem, which conserves the total energy.
They predict shrinkage of the photospheric radius whenever the magnetic energy is increased, so that a minimum solar radius
should occur in the maximum phase of the solar activity. If the magnetic flux tubes are produced near the base of the envelope:
$M \approx 10^{32}$ $Joules$ \cite{Spie80}. If it is generated in the super-adiabatic region near the surface:
$M \approx 10^{28}$ $Joules$  \cite{Dear82}. By comparison, we have $|\Omega| \approx 10^{42}$ $Joules$. Moreover, within the
solar convective zone, $\zeta$ must be very small, but it is not yet clear how small it must be.

\section{What Is the Shape of the Sun?}
If the Sun is described as a sphere, then the gravitational and the pressure gradient forces are in hydrodynamic equilibrium.
But due to the non-homogenous mass distribution and the differential rotation inside the Sun, its outer shape turns out to be
distorted in latitude. The Sun has an extended atmosphere and it is not so simple to consider the upper limit of its photosphere.
\cite{Roze03} have defined the free surface of the Sun as a level, where a given physical parameter such as the temperature,
the density, the pressure, etc. is constant. This free surface does not exactly coincide with an ellipsoid of revolution.
The true figure is a bit far more complex showing 'asphericities' (measured by 'shape coefficients'); to first order it lies
between the spherical and ellipsoid shape, which is a figure called 'spheroid'; this could be recognized by the flattening
parameter, $f$ (or by the oblateness, $\varepsilon = \frac{R_{eq} - R_{pol}}{R_{sol}}$). If the Sun is considered as an
ellipsoid of equatorial ($R_{eq}$) and polar ($R_{pol}$ radii, the solar radius (limited to order $2$) can be written (after some reductions) as,
\begin{equation}
r= R_{eq} (1- \frac{1}{3} f - \frac{2}{3} fP_{2}),
\end{equation}
where $f=\frac{R_{eq} - R_{pol}}{R_{sp}}$, is the flattening; $R_{sp}= (R_{eq}^2 R_{pol})^{\frac{1}{3}}$, the radius of the
best sphere passing through the equatorial and the polar radii which is determined by applying $P_{2}=0$; and
$P_{2}= P_{2}(\cos \theta)$, the Legendre Polynomial. Thus, as $R_{sp} =R_{eq}(1- \frac{1}{3} f$, one gets
\begin{equation}
\frac{1}{r}= \frac{1}{R_{sp}} (1+ \frac{2}{3} fP_{2}) + O(f^2),
\end{equation}
From this formalism, we will calculate in the next section, the gravitational and the magnetic energy for two cases, a spherical and an oblate Sun.

\section{Calculation}

By considering the variability of the magnetic energy, we derive the magnetic field and the magnetic energy by a suitable
vector potential. The magnetic field in the tachocline (a layer between the radiation and the convection regions)
is toroidal, so we have $A=(0,0,A_{\varphi})$ and $\textbf{B}= \nabla $$\times$ $\textbf{A}_{\varphi}$ in a spherical coordinate
system. The magnetic field is described by a polynomial function \cite{Dur97}:
\begin{equation}
B_{r}= \sum c_{l}(2l+1)^{\frac{1}{2}} P_{l}(\cos \theta),
\end{equation}
where
\begin{equation}
c_{l} = \frac{1}{2} \int_{0}^{\pi} B_{r}(2l+1)^{\frac{1}{2}}P_{l}(\cos \theta) \sin(\theta) d\theta.
\end{equation}
To derive a variability of the gravitational potential energy, we write:
\begin{equation}
\Phi_{g}(r, \theta, \varphi) = \sum_{n=0}^{\infty} \frac{1}{r^{n+1}} \sum_{m=0}^{\infty} [A_{nm}P_{nm}(\cos \theta)
\cos (m\varphi) + B_{nm} P_{nm}(\cos \theta) \sin(m \varphi)],
\end{equation}
where $r$ is the radius and $P_{nm}$ is the Legendre polynomial of degree $n$ and order $m$. We suppose that the rotation is symmetric, so that
\begin{equation}
\Phi_{g}(r, \theta, \varphi) = -\frac{GM}{r} [1- \sum_{n=1}^{\infty} (\frac{a}{r})^{n} J_{n} P_{n}(\cos \theta)].
\end{equation}
Here $J_{n}$ is the gravitational moment of order $n$. Helioseismology data imply a significant temporal variation on both the angular 
momentum and the gravitational multipolar moments. Using such temporal variations would permit to constrain dynamical theories of the 
solar cycle. As $J_{n}$ are not of enough magnitude for $n > 4$, we will use here the order $2$ only, then \cite{Faz07}
\begin{equation}
\Phi_{g} = -\frac{GM}{r} [1- (\frac{a}{r})^{2} J_{2} P{2}(\cos \theta)].
\end{equation}
From the above considerations, the two potentials can be written as:
\begin{equation}
\Phi_{g} = -\frac{GM}{R_{sp}} [1+ \frac{2}{3} f - J_{2} P_{2}],
\end{equation}
\begin{equation}
A_{\varphi} = - (\frac{9\surd2}{8} B_{cr}) R_{sp}[1- \frac{2}{3} f P_{2}].
\end{equation}
The magnetic energy, $M = \int\frac{B^{2}}{2\mu_{0}} dV$, and the gravitational energy, $\Omega = \frac{1}{2} \int\rho\Psi dV$, 
are calculated as followed: For a spherical Sun,
\begin{equation}
M_{\textit{spherical}} = \int \frac{|(\nabla\times A_{\varphi})^{2}|}{2\mu_{0}} dV = - \frac{9\surd2}{8}B_{cr}\int
\frac{r}{2\mu_{0}} dV,
\end{equation}
\begin{equation}
\Omega_{\textit{spherical}} = \int\rho\frac{-GM}{r} dV = - GM \int\frac{\rho}{r} dV.
\end{equation}
And for a spheroidal Sun,
\begin{equation}
M_{\textit{spheroid}} = \int \frac{|(\nabla\times A_{\varphi})^{2}|}{2\mu_{0}} dV = - \frac{9\surd2}{8}B_{cr}\int
\frac{r_{sp}(1-\frac{2}{3} f P_{2})}{2\mu_{0}} dV,
\end{equation}
\begin{equation}
\Omega_{\textit{spheroid}} = \int\rho\frac{-GM}{r} [1+ \frac{2}{3} f - J_{2} P_{2}]dV = - GM \int\frac{\rho}{r_{sp}} [1+ \frac{2}{3} f - J_{2} P_{2}]dV.
\end{equation}

\section{Results}
We compare the magnetic and the gravitational energies for the considered shapes of the Sun. Figure 1 shows the variation of the magnetic
energy as a function of the solar latitude for a spherical (upper) and a spheroidal (lower) shapes. For upper plot, the maximum occurs at the
solar equator, and then decreases toward the pole (for sake of clarity, only two curves are displayed, one for $r = 0.69R$ and the other one
for $r = 0.73R$, as indicated in the left box). For lower plot, the variation of the magnetic energy has a local minimum at $28^{\circ}$ and then
increases to values less than the equator one (three curves are displayed, for $r = 0.69R$, $r = 0.71R$ and $r = 0.73R$, as indicated in the right box).
\begin{figure}
 \centerline{\includegraphics[width=5cm]{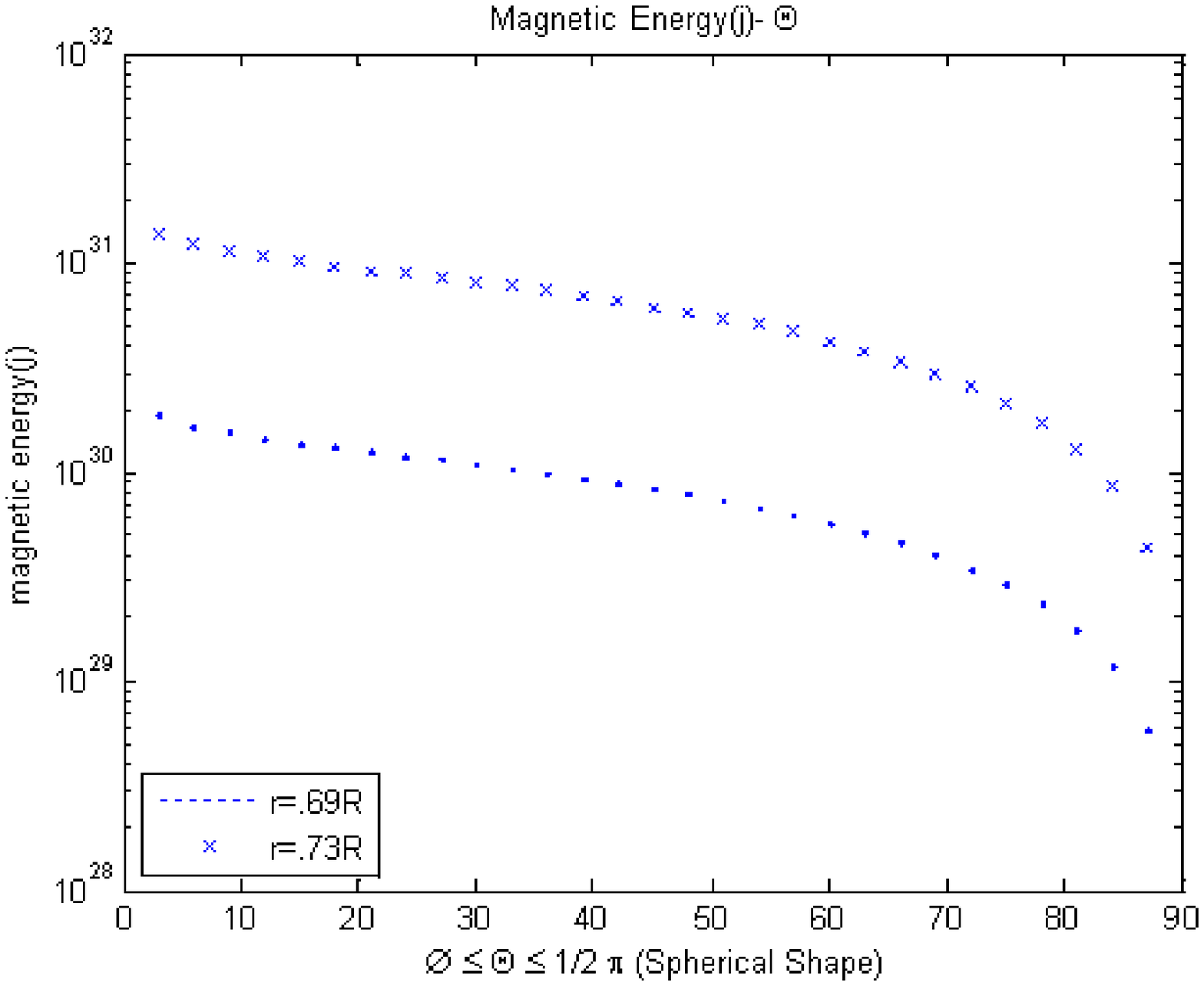}}
 \centerline{\includegraphics[width=5cm]{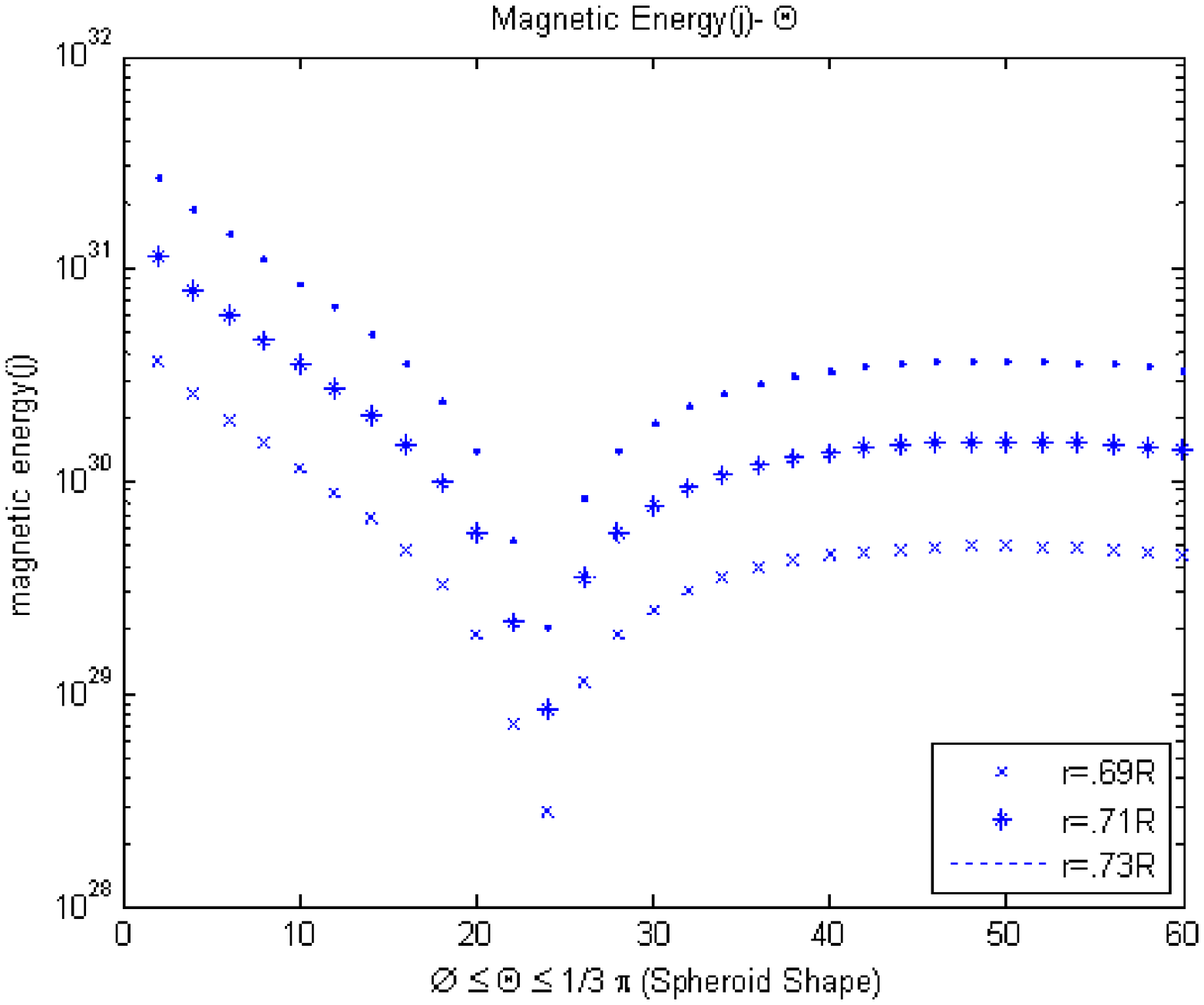}}
 \caption[]{Variation of the magnetic energy with respect to the solar latitude for two different radii
 for spherical shape (upper) and spheroid shape (lower). }
 \label{fig1}
\end{figure}
\\Fig. 2 displays the variation of magnetic energy as a function of solar latitude for $r = 0.72R$, i.e.,
at the center of the tachocline (the upper curve being the spherical shape and the lower one the spheroidal shape).
Finally, in Fig. 3 we indicate the variation of the magnetic energy as a function of radius, at the polar latitude
($90^{\circ}$) for the considered solar shapes. For sake of clarity, the upper curve has been shifted one order of magnitude up. 	
\begin{figure} 
 \centerline{\includegraphics[width=5cm]{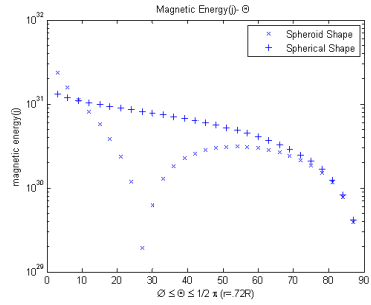}}
 \caption[]{Variation of the magnetic energy as a function of solar latitude only for $r = 0.72R$ in two types of the solar shape.}
 \label{fig2}
\end{figure}
\begin{figure} 
 \centerline{\includegraphics[width=5cm]{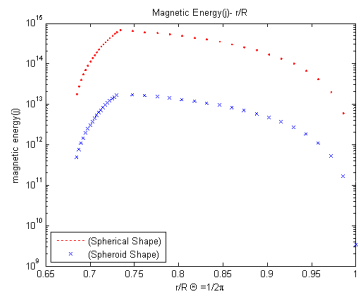}}
 \caption[]{Variation of the magnetic energy as a function of solar radius at the polar latitude ($90^{\circ}$) in two types of the solar shape.}
 \label{fig3}
\end{figure}
\\Fig. 4 shows variation of the gravitational energy as a function of the solar radius, for three different heliographic latitudes ($0^{\circ}$,
$30^{\circ}$ and $60^{\circ}$, as shown in the right boxes) for two cases of solar shape (which are mentioned in figures), respectively.
\begin{figure} 
 \centerline{\includegraphics[width=5cm]{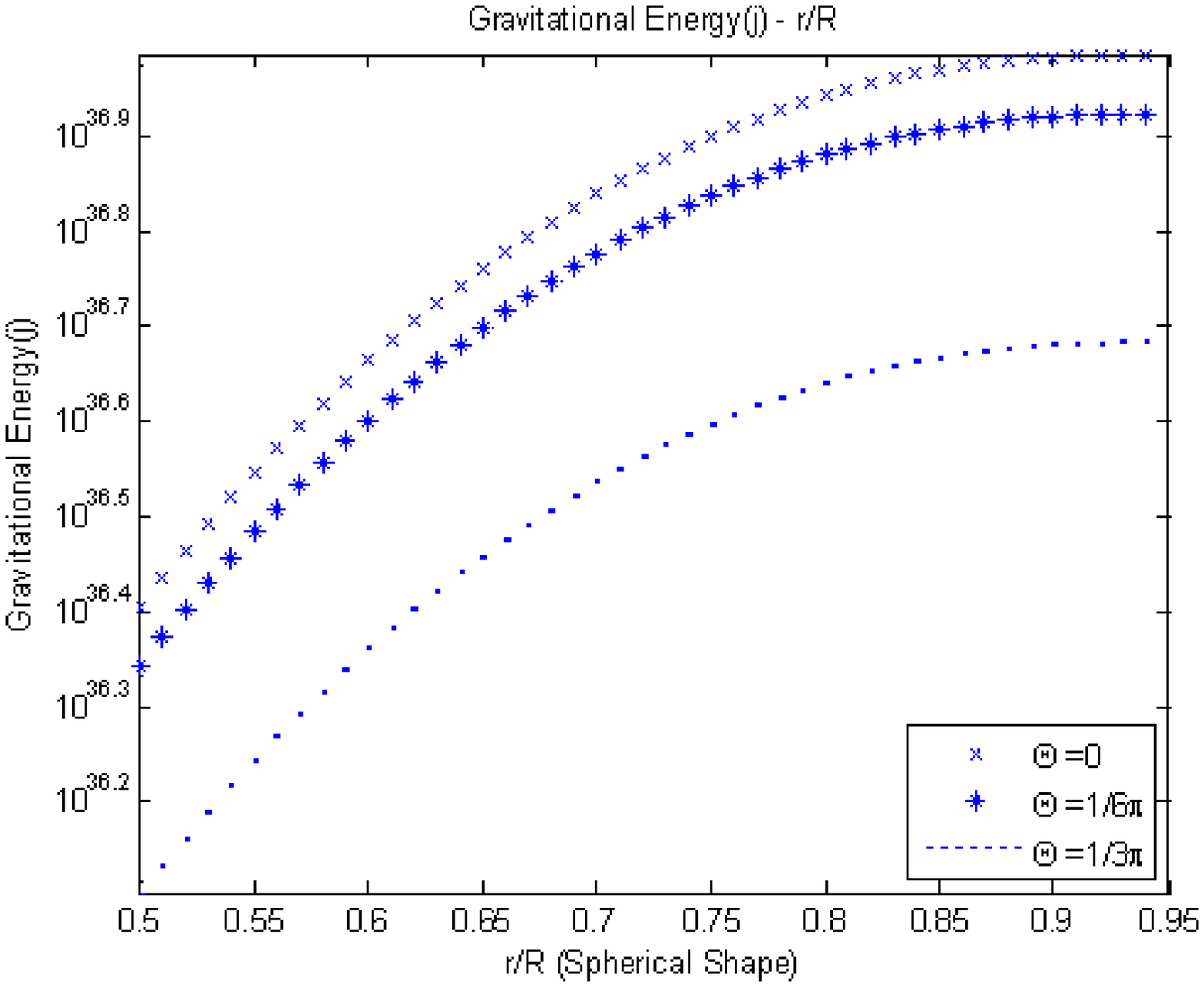}}
 \centerline{\includegraphics[width=5cm]{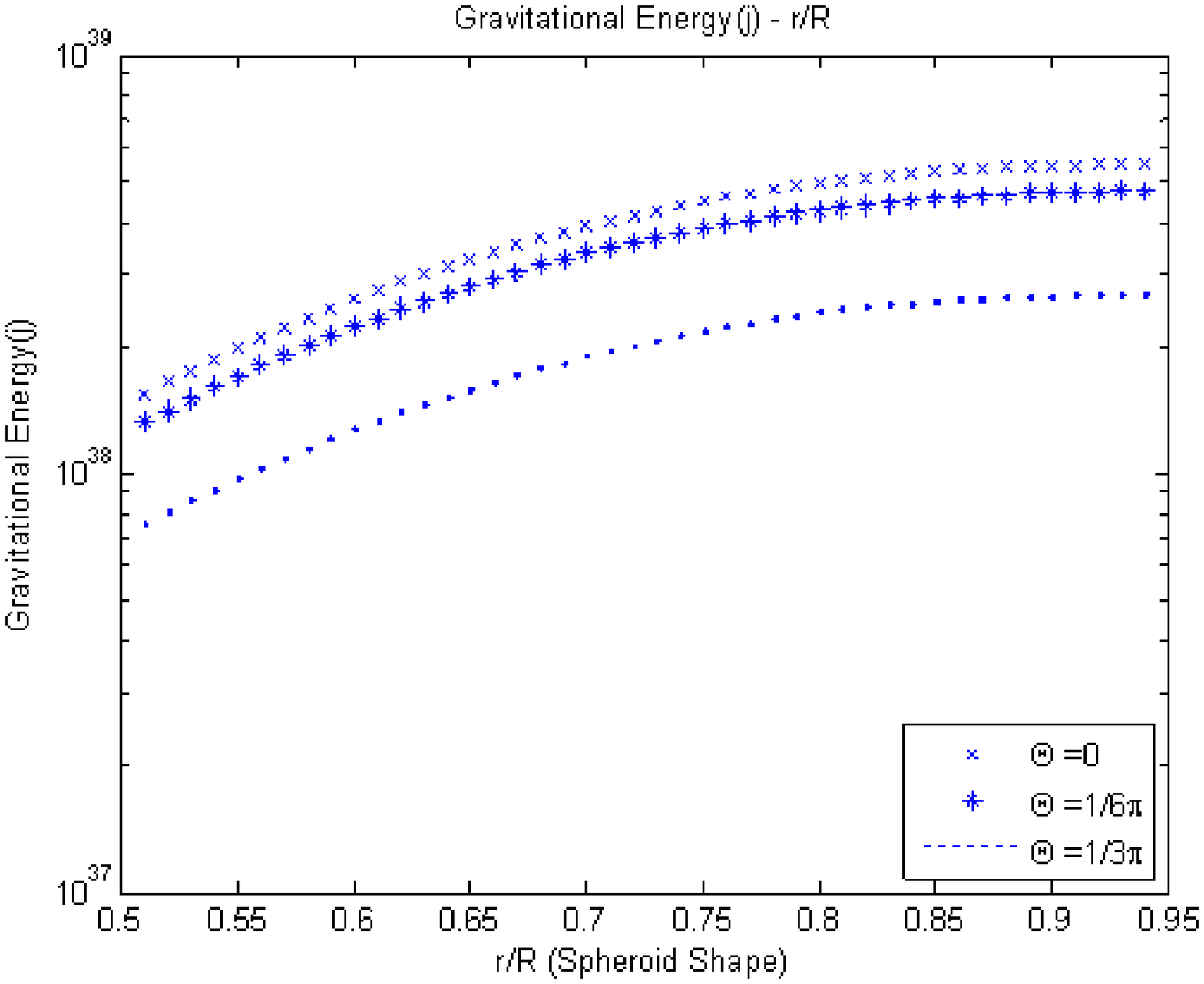}}
 \caption[]{variation of the gravitational energy as a function of the solar radius, for three different solar latitudes 
 in two cases of solar shape (which are mentioned in figures). }
 \label{fig4}
\end{figure}
\\In Fig. 4 (lower), a similar variation of the gravitational energy has been obtained for three considered latitudes from the base of 
the convective zone up to $0.95R$. By increasing the magnetic field in the tachocline layer, the magnetic energy is also increasing, 
while by rising from the tachocline to the surface, the magnetic energy decreases along with the magnetic field.
\\Comparing Figs. 3 and 4, the magnetic energy is decreasing from the tachocline to the solar surface, while the gravitational energy 
is gradually increased. It seems that these opposite variations of magnetic and gravitational energies are related to the distribution 
of density in the solar interior: within the tachocline where the density is slightly higher than its general decrease trend, the 
gravitational energy increases. By decreasing of density from the tachocline to the surface, the energy variation is smoother in 
magnitude (Fig. 4, lower). Moreover, the variation of the magnetic energy stored in the magnetic field (Fig. 3) depends on the 
magnetic field's behavior with the solar density (when the density decreases, so does the current and the magnetic field).
Finally, Figs. 5 and 6 show the variation of the gravitational energy as a function of the solar radius (at $60^{\circ}$ of latitude) 
and as a function of the solar latitude (at $r = 0.7R$) for two spherical and spheroid shapes, respectively. Here also, for sake of clarity, 
the upper curve has been shift up of one order of magnitude.
\begin{figure} 
 \centerline{\includegraphics[width=5cm]{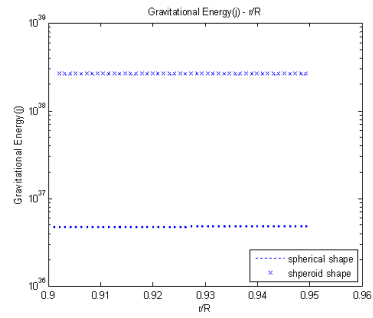}}
 \caption[]{variation of the gravitational energy as a function of the solar radius at $60^{\circ}$ of latitude.}
 \label{fig5}
\end{figure}
\begin{figure} 
 \centerline{\includegraphics[width=5cm]{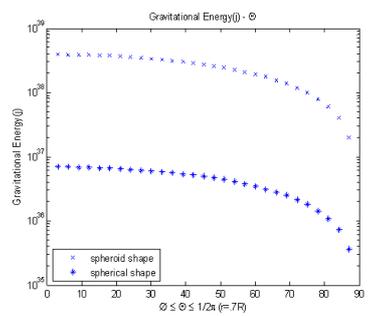}}
 \caption[]{variation of the gravitational energy as a function of the solar latitude at $r = 0.7R$.}
 \label{fig6}
\end{figure}

\section{Discussion}

In this paper, we considered radially variations of the magnetic field and we assumed that there exist a relation between 
distribution of the magnetic field and the differential rotation in the solar convective zone. Our results are investigated 
for two types of the solar shape: spherical and spheroidal shapes. \cite{Endal85} proposed that a variable internal magnetic 
field should affect all global parameters of the Sun such as radius and irradiance. \cite{Lef03} found distortions on the 
solar shape, i.e. a faint bump near $30^{\circ}$ to $50^{\circ}$. Here, we obtained an increase of the magnetic energy 
from about $28^{\circ}$ up to $50^{\circ}$ (with a small percentage of error $\approx 3^{\circ}$). So we argue that these 
distortions may be connected to this energy variation. But the magnetic energy variation is thwarted by a decrease of the 
differential rotation above the tachocline layer. We found that the magnetic energy undergoes a decreasing steep from 
$50^{\circ}$ to $90^{\circ}$ which agrees to the decreasing differential rotation towards the pole.

Variation of the magnetic field in the spherical shape is correlated with the distribution of the differential rotation (Fig. 1, upper). 
When differential rotation decreased at high latitudes, the magnetic field must be decreased. This diminishing undergoes until the 
magnetic energy becomes minimum at the pole, which is similar to the behavior of the magnetic field in the solar spheroidal shape.

Fig. 1 indicates that: a) for the spherical shape, the magnetic energy variation has a minimum at the poles and a maximum at the 
equator but, b) for the spheroidal shape, there is a maximum at the equator with a fast decreasing around $28^{\circ}$; another 
maximum takes place at $50^{\circ}$ and then the magnetic energy decreases up to the pole. The maximum value of magnetic energy 
at both equator and pole reach to $10^{32}$$J$ which is in agreement with the observed solar irradiance variations in each solar 
cycle. By contrast, the gravitational energy shows the same variation for two types of the solar shapes (Fig. 4).

Since the maximum of magnetic energy variations is stored in the toroidal magnetic field at the equator at the bottom of the 
convection zone, it would be considered as a source for generation of the poloidal component of the magnetic field. This 
component is weak at the equator and increases toward the poles. After the generation phase of the poloidal field, latitudinal 
differential rotation of the Sun begins to stretch the field lines beneath the faster-spinning equatorial regions. The poloidal 
field (north-south) is essentially changed into a toroidal (east-west) configuration by the meridional motions, in which the 
lines of force are near-circles around the solar axis. So the toroidal field must be produced at the poles and when reaching 
the equator, it becomes much stronger. The stretching process wraps the lines around the Sun several times causing them to 
intertwine and intensify; finally they can be driven to the surface as magnetic loops \cite{Dur97}

Study of variations of the magnetic and the gravitational energies is important to develop the solar dynamo theory. 
We showed that regions of strong sub-surface rotational shear allow the toroidal magnetic flux to penetrate to the 
solar surface. In other words, the magnetic toroidal field generated in the sub-surface shear layer (the leptocline) 
does not contribute directly to the generation of the poloidal magnetic field; a fact that already recognized by \cite{Pip11}, 
the subsurface-shear layer shaped the solar $\alpha\Omega$ dynamo.


\end{document}